\documentclass[aps,epsf,nofootinbib,
notitlepage,showpacs,10pt, unsortedaddress, superscriptaddress]{revtex4-2}

\usepackage[svgnames, table]{xcolor}

\usepackage{color}
\usepackage{float}
\usepackage{graphicx}
\usepackage{subcaption}
\usepackage{amsmath}
\usepackage{amsthm}
\usepackage{mathtools}
\newtheorem{theorem}{Theorem}
\theoremstyle{definition}

\usepackage{bm}
\usepackage{hyperref}
\usepackage{float}
\usepackage{multirow}
\restylefloat{table}
\usepackage{caption}

\newcommand{\bra}[1]{{\left\langle #1 \right|}}
\newcommand{\ket}[1]{{\left| #1 \right\rangle}}

\theoremstyle{definition}

\usepackage{tikz} 
\usetikzlibrary{matrix,positioning}
\usepackage{amsfonts}
\usepackage{amssymb}
\newcommand{\rh}[1]{{\textcolor{blue}{#1}}} 
\newcommand{\moi}[1]{{\textcolor{magenta}{#1}}}

\hypersetup{
   colorlinks = true,
    urlcolor=blue
}
\usepackage[normalem]{ulem}
\usepackage{lipsum}

\graphicspath{{figures/}}
\usepackage{tikz-network}
\usetikzlibrary{fit, shapes.arrows}
\usetikzlibrary{shapes.geometric, arrows}
\usetikzlibrary{positioning,chains,fit,shapes,calc}

\tikzstyle{boxblue} = [rectangle, rounded corners, minimum width=3cm, minimum height=1cm,text centered, draw=black, fill=blue!30]
\tikzstyle{boxred} = [rectangle, rounded corners, minimum width=3cm, minimum height=1cm,text centered, draw=black, fill=red!30]
\tikzstyle{boxgreen} = [rectangle, rounded corners, minimum width=3cm, minimum height=1cm,text centered, draw=black, fill=green!30]
\tikzstyle{io} = [trapezium, trapezium left angle=70, trapezium right angle=110, minimum width=3cm, minimum height=1cm, text centered, draw=black, fill=orange!30]
\tikzstyle{process} = [rectangle, minimum width=3cm, minimum height=1cm, text centered, draw=black, fill=orange!30]
\tikzstyle{decision} = [diamond, minimum width=3cm, minimum height=1cm, text centered, draw=black, fill=green!30]
\tikzstyle{arrow} = [thick,->,>=stealth]

\tikzstyle{decision} = [diamond, draw, fill=blue!20, 
    text width=4.5em, text badly centered, node distance=3cm, inner sep=0pt]
\tikzstyle{block} = [rectangle, draw, fill=blue!20, 
    text width=5em, text centered, rounded corners, minimum height=4em]
\tikzstyle{line} = [draw, -latex']
\tikzstyle{cloud} = [draw, ellipse,fill=red!20, node distance=3cm,
    minimum height=2em]

\usepackage{float}

\usepackage{array, booktabs, makecell, multirow}

\usepackage{amsmath,calc}
\newlength{\LPlhbox}

\newcommand\setItemnumber[1]{\setcounter{enumi}{\numexpr#1-1\relax}}

\definecolor{graphPurple}{RGB}{185,75,185}

\usepackage{algorithm}
\usepackage{algpseudocode}

\usepackage[normalem]{ulem}
\usepackage{subcaption}
\usepackage{caption}
\usepackage{bm}
\algnewcommand{\To}{\textbf{To }}
\algnewcommand\Input{\item[\textbf{Input:}]}%
\algnewcommand\Output{\item[\textbf{Output:}]}%

\begin{document}

\markboth{First author}
{Graph decomposition techniques for solving CO problems with VQAs}

\title{Graph decomposition techniques for solving combinatorial optimization problems with variational quantum algorithms}

\author{Moises Ponce}
\email{mponce@tennessee.edu}
\affiliation{
	Department of Industrial and Systems Engineering\\ University of Tennessee at Knoxville\\Knoxville, TN 37996}

\author{Rebekah Herrman}
\affiliation{
	Department of Industrial and Systems Engineering\\ University of Tennessee at Knoxville\\Knoxville, TN 37996}
	
	\author{Phillip C. Lotshaw}
\affiliation{
	Quantum Computational Science Group\\ Oak Ridge National Laboratory\\ Oak Ridge, TN 37830} \thanks{This manuscript has been authored by UT-Battelle, LLC, under Contract No. DE-AC0500OR22725 with the U.S. Department of Energy. The United States Government retains and the publisher, by accepting the article for publication, acknowledges that the United States Government retains a non-exclusive, paid-up, irrevocable, world-wide license to publish or reproduce the published form of this manuscript, or allow others to do so, for the United States Government purposes. The Department of Energy will provide public access to these results of federally sponsored research in accordance with the DOE Public Access Plan.}
\author{Sarah Powers}
\affiliation{
	Computer Science and Mathematics Division\\ Oak Ridge National Laboratory\\ Oak Ridge, TN 37830}

\author{George Siopsis}
\affiliation{
	Department of Physics and Astronomy\\ University of Tennessee at Knoxville\\Knoxville, TN  37996-1200}

	\author{Travis Humble}
\affiliation{
	Quantum Science Center\\ Oak Ridge National Laboratory\\ Oak Ridge, TN 37830}
 
\author{James Ostrowski}
\email{jostrows@tennessee.edu}\thanks{corresponding author}
\affiliation{
	Department of Industrial and Systems Engineering\\ University of Tennessee at Knoxville\\Knoxville, TN 37996}

\begin{abstract}
    The quantum approximate optimization algorithm (QAOA) has the potential to approximately solve complex combinatorial optimization problems in polynomial time. However, current noisy quantum devices cannot solve large problems due to hardware constraints.  In this work, we develop an algorithm that decomposes the QAOA input problem graph into a smaller problem and solves MaxCut using QAOA on the reduced graph. The algorithm requires a subroutine that can be classical or quantum--in this work, we implement the algorithm twice on each graph. One implementation uses the classical solver Gurobi in the subroutine and the other uses QAOA. 
    We solve these reduced problems with QAOA. On average, the reduced problems require only approximately 1/10 of the number of vertices than the original MaxCut instances. 
    Furthermore, the average approximation ratio of the original MaxCut problems is 0.75, while the approximation ratios of the decomposed graphs are on average of 0.96 for both Gurobi and QAOA. 
    With this decomposition, we are able to measure optimal solutions for ten 100-vertex graphs by running single-layer QAOA circuits on the Quantinuum trapped-ion quantum computer H1-1, sampling each circuit only 500 times. 
    This approach is best suited for sparse, particularly $k$-regular graphs, as $k$-regular graphs on $n$ vertices can be decomposed into a graph with at most $\frac{nk}{k+1}$ vertices in polynomial time. 
    Further reductions can be obtained with a potential trade-off in computational time. 
     While this paper applies the decomposition method to the MaxCut problem, it can be applied to more general classes of combinatorial optimization problems. 
\end{abstract}

\maketitle

\section{Introduction}

Current quantum devices, commonly referred to as noisy intermediate scale quantum (NISQ) devices, are limited by lack of error correction and imperfect gate control, so shallow circuit depth implementations of algorithms are a necessity \cite{Preskill2018NISQ, Alexeev_2021QCSytemsforSCDisc,Bharti_2022}. Despite the limitations in this regime, a great deal of success has been witnessed in utilizing hybrid quantum-classical algorithms. Variational quantum algorithms (VQAs) utilize a quantum computer to perform operations that require real-valued parameters on some initial state. An outer loop optimizes the parameters via classical optimization techniques \cite{McClean_2016TheoryVHybrid,Cerezo_2021VQAs}.

The quantum approximate optimization algorithm (QAOA) \cite{Hogg2000QAOA,Farhi2014FQAOA} is a promising candidate for quantum primacy and is universal for computation \cite{FarhiHarrow2016QAOASup, Morales_2020CompUniversalityQAOA,lloyd2018quantum, herrman2022relating}. Since it is a VQA, it requires classical parameter input, and the parameter choice greatly affects the solution output by the algorithm. There are several classical optimization techniques used to find the optimal parameters, such as reinforcement learning, numerical optimization algorithms, and heuristics \cite{wauters2020reinforcement, Zhou_2020QAOAImplementationPerformance,shaydulin2022transfer,lotshaw2021empirical}. Typically, larger problems require deeper circuit depth, which may introduce prohibitive amounts of noise in near-term processors and also complicates parameter optimization \cite{Guerreschi_2019,Willsch_2020,Zhou_2020QAOAImplementationPerformance, Shaydulin_2019MultiStartQAOA,lotshaw2022scaling,lotshaw2022approximate}. QAOA variations have been introduced that incorporate additional classical parameters and perform at least as well as the original algorithm, even strictly better in some cases \cite{herrman2022multi, vijendran2023expressive, wurtz2021classically}. Previous QAOA research indicates that problem structure impacts solution quality and parameter choice \cite{farhi2020quantum, herrman2021impact, shi2022multi, shaydulin2020classical}. 


 Classical optimization methods such as dualization and preprocessing have been studied in the context of QAOA, both in terms of initial state preparation and solution quality \cite{herrman2021globally, tate2020bridging}. More closely related to the current work, two recent approaches, QAOA-in-QAOA (denoted $\mathrm{QAOA}^2$) and multilevel quantum local search (ML-QLS) consider iterative approaches to approximately solving large combinatorial problems by solving sequences of subproblems \cite{ushijima2021multilevel,zhou2023qaoa}. 
 The solutions to the subgraphs are merged together, and QAOA is used to solve the resulting modified graph. The technique introduced in this paper differs from them because it solves subgraphs of the original MaxCut graph in series. For three-regular graphs, the solved subgraphs are used to create a modified MaxCut graph that has the same objective as the original graph. 
 It is unclear, however, how other methods used to solve combinatorial optimization (CO) problems in a classical regime can be used to enhance VQAs. 
 
 In this work, we demonstrate that implementing a decomposition technique on a graph and then running QAOA on the resulting simplified, often weighted, graph can output a solution comparable to, and possibly better than, the original algorithm. Furthermore, in many cases, the number of qubits required to solve the decomposed problem is significantly fewer than the number required to solve the original. 
 The decomposition algorithm relies on calculating cut sets of graphs, which can be found in polynomial time, and solving integer programs on small instances to create a reweighted graph, which can be done using classical techniques. Furthermore, the decomposition method can be used on general CO problems that can be formulated as graphs. 

The rest of the paper is organized in the following manner: In Section~\ref{sec:background}, we introduce terminology on graphs, QAOA, and combinatorial optimization problems. In Section~\ref{sec:graphdecomp}, we describe our decomposition approach on QUBO problem, outline an algorithmic framework, display case on applying the decomposition technique on a MaxCut instance, and provide an example of the decomposition algorithm iterating on a single graph. We then discuss the effect of decomposition techniques as preprocessing for QAOA in Section~\ref{sec:qaoadecomp}. Finally, we conclude with a discussion in Section~\ref{sec:discussion}. 


\section{Background}\label{sec:background}
We now introduce relevant graph theory, QAOA, and CO problem terminology and concepts.
\subsection{ Graph Theory}\label{subsec:graphbackground}

An undirected weighted graph $G=(V,E,J)$ consists of three sets: \textit{vertices}, $V$, \textit{edges}, $E$, and \textit{weights}, $J$. $V$ is a nonempty, typically finite set, while the set $E$ is a collection of unordered pairs in $V \times V$. The $|V| \times |V|$ matrix $J$ encodes the weights of the graph, where $J_{ij}$ contains the weight of edge $(i,j),$ for $i$ not equal to $j$, and $J_{ii}$ contains the weight of vertex $i$. 

\begin{figure}
    \centering
    \begin{subfigure}{0.3\textwidth}
        \begin{tikzpicture}[scale = 0.6]
        \Vertex [IdAsLabel, y=1] {1} \Vertex[IdAsLabel,x=3,y=3]{2} \Vertex[IdAsLabel, x=6, y =1]{3} \Vertex[IdAsLabel, x=3,y=-1]{4}
        \Vertex[IdAsLabel, x=8, y = 3]{5} 
        \Edge(1)(2)
        \Edge(1)(3)
        \Edge(1)(4)
        \Edge(4)(3)
        \Edge(3)(5)
        \end{tikzpicture}
    \end{subfigure}%
    \begin{subfigure}{0.3\textwidth}
        \begin{tikzpicture}[scale = 0.6]
        \Vertex [IdAsLabel, y=1, ,RGB,color={190,174,212}] {1} \Vertex[IdAsLabel,x=3,y=3, ,RGB,color={190,174,212}]{2} \Vertex[IdAsLabel,opacity =.2, x=6, y =1]{3} \Vertex[IdAsLabel, x=3,y=-1,RGB,color={190,174,212}]{4}
        \Vertex[IdAsLabel, x=8, y = 3,RGB,color={45,180,25} ]{5} 
        \Edge(1)(2)
        \Edge(1)(4)
        \end{tikzpicture}
    \end{subfigure}
    \begin{subfigure}{0.3\textwidth}
        \begin{tikzpicture}[scale = 0.6]
        \Vertex [IdAsLabel, y=1,RGB,color={255,30,115}] {1} \Vertex[IdAsLabel,x=3,y=3,RGB,color={255,175,30}]{2} \Vertex[IdAsLabel, x=6, y =1, ,RGB,color={255,175,30}]{3} \Vertex[IdAsLabel, x=3,y=-1, ,RGB,color={255,175,30}]{4}
        \Vertex[IdAsLabel, x=8, y = 3]{5} 
        \Edge(1)(2)
        \Edge(1)(3)
        \Edge(1)(4)
        \Edge(4)(3)
        \Edge(3)(5)
        \end{tikzpicture}
    \end{subfigure}
    \caption{Left: An example of a graph. Middle: the minimum cut set $K = \{3\}$ that splits the graph into two nonempty components, purple and green. Right: The neighbors of vertex 1 are colored in orange.}
    \label{fig:example1}
\end{figure}
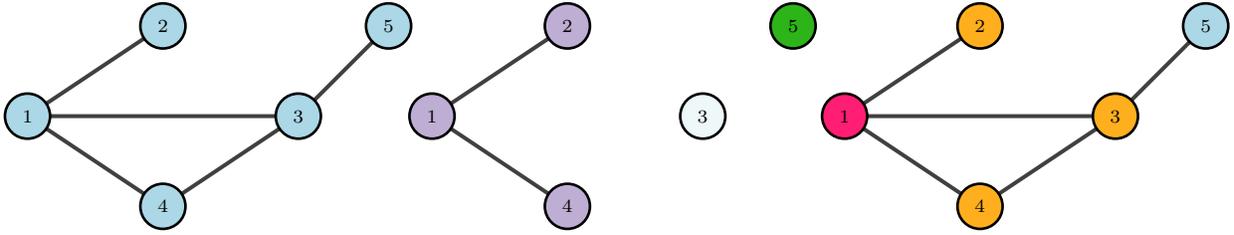

A graph is said to be \textit{connected} if one can traverse edges of the graph from one vertex to any other vertex in the graph. For a given connected graph, the removal of a set of vertices can disconnect into a number of components. A set of vertices whose removal disconnects a graph is called a \textit{vertex cut set}. For example, it is possible to split the graph in Fig.~\ref{fig:example1} into three disconnected components by removing vertex $1$ and $3$. A \textit{minimum vertex cut set} is the minimum number of vertices whose removal disconnects the graph into at least two nonempty components. The minimum number of vertices needed to disconnect the graph in Fig.~\ref{fig:example1} is one: vertex $3$ disconnects the graph on its own, as does vertex $1$. Let $S \subset V(G)$. The \textit{induced subgraph}, $G[S]$, is a graph that has vertex set $S$ and its edge set  consists of all edges in $G$ that have both endpoints in $S$. 
A graph is said to be $k$-regular if every vertex has $k$ neighbors. Fig.~\ref{fig:example1} is not $k$-regular. 

The aim of the MaxCut problem is to partition the vertex set of a graph into two disjoint sets such that the number of edges with endpoints in each set is maximized. One application of MaxCut is finding the ground states of quantum spin models such as spin glasses \cite{de1995exact, liers2004computing,lotshaw2023simulations}
\subsection{QAOA}\label{subsec:qaoabackground}

QAOA is a classical-quantum hybrid algorithm that requires real-valued parameters input into a parameterized quantum circuit. It requires two unitary operators
\begin{equation*}
U(\beta) = e^{-i \beta H_{B}}
\end{equation*} and
\begin{equation*}
U(\gamma) = e^{-i \gamma H_{C}},
\end{equation*}
\noindent where $H_B$ is called a mixer Hamiltonian, $H_C$ is the cost Hamiltonian that encodes the CO problem being solved, and $\gamma$ and $\beta$ are both real valued numbers between $0$ and $2\pi$.   
 The unitary operators are applied to an initial state $\ket{s}$, which is an eigenstate of $H_B$, in an alternating fashion $p$ times to obtain a final state $\ket{\boldsymbol{\gamma}, \boldsymbol{\beta}}$, 

\begin{equation*}
\ket{\boldsymbol{\gamma}, \boldsymbol{\beta}} = e^{-i \beta_{p}H_{B}} e^{-i\gamma_{p}H_{C}} \ldots e^{-i \beta_{1}H_{B}} e^{-i\gamma_{1}H_{C}} \ket{s}
\end{equation*}


\noindent This state is then measured to obtain the solution to the CO problem being solved. 

The expected value of the cost Hamiltonian is denoted $\langle H_C \rangle$ and is equal to 
\begin{equation*}
\langle H_C \rangle = \bra{\psi(\boldsymbol{\gamma}, \boldsymbol{\beta})}H_C\ket{\psi(\boldsymbol{\gamma}, \boldsymbol{\beta})}.
\end{equation*}

\noindent The primary metric for success for QAOA is the \textit{approximation ratio}, abbreviated $A.R.$, which is

\begin{equation*}
    A.R. = \frac{\langle H_C\rangle}{C_{max}},
\end{equation*}
\noindent where $C_{max}$ is the optimal solution to the CO problem being solved. The value of $\langle H_C \rangle$ relies on the choice of $\gamma$ and $\beta$, and since the approximation ratio is the primary success metric, $\gamma$ and $\beta$ are chosen to maximize $\langle H_C \rangle$. 

\subsection{Combinatorial optimization problems}\label{subsec:cobackground}


The goal of combinatorial optimization problems is to maximize or minimize some objective function  
\begin{equation*}
C(z) = \sum_{a} C_a(z)
\end{equation*}
\noindent where $z$ is a bit string of length $n$, and each $C_a(z)$ in the sum is referred to as a clause. In order to solve these types of problems with QAOA, the classical clauses $C_a(z)$ must be converted into Hamiltonians. See \cite{hadfield2021representation} for a detailed overview of how Booleans can be mapped into Hamiltonians.



The MaxCut problem \cite{Karp1972} is an NP-complete combinatorial optimization problem that has been the focus of recent QAOA research. 
In order to map the classical MaxCut formulation to a Hamiltonian, we assign a qubit to each $v \in V$. Each clause in the classical formulation is written in terms of an edge of the graph, $e_{i,j}$, and is written as

\begin{equation*}
C_{i,j} =  \frac{1}{2}  ( \mathbb{I} - \sigma_{i}^{z}\sigma_{j}^{z}),
\end{equation*}
\noindent in the Hamiltonian formulation. The total cost Hamiltonian is the sum of clauses
\begin{equation*}
    H_C = \displaystyle\sum_{ e_{i,j} } C_{ i,j }
\end{equation*}
\noindent where $\sigma_{i}^{z}$ refers to the Pauli-z operator acting on qubit $i$, $\mathbb{I}$ is the $2^n$ identity matrix, and $|V| = n$.

There are various classical approximate algorithms for solving the MaxCut problem such as Goemans-Williamson's (GW) algorithm \cite{goemans1995improved}, which is guaranteed to achieve an approximation ratio of $0.879$.
However, QAOA is guaranteed to find the optimal solution as the number of iterations $p \rightarrow \infty$ \cite{Farhi2014FQAOA}. This work will focus on using decomposition techniques to solve the MaxCut problem.


\section{Graph Decomposition on QUBO Problems}\label{sec:graphdecomp}



Quadratic unconstrained binary optimization (QUBO) problems are one type of CO problem. In a QUBO, the objective function is a quadratic function over a set of binary variables,

\begin{equation*}
C(z) = \sum_{i} \sum_{j\neq i} J_{ij}z_i z_j  + \sum_{i} J_{ii} z_i. \label{eq:objctive}
\end{equation*}

\noindent For these types of problems, the goal is to find a bitstring $z^*$ that either maximizes or minimizes $C(z)$, that is to find
\begin{align}
C_{max} = \max_{z \in \{0,1\}^n} \sum_{i} \sum_{j\neq i} J_{ij}z_i z_j  + \sum_{i} J_{ii} z_i. \label{eq:qubo}.
\end{align}


\noindent 

Visualize~\eqref{eq:qubo} as a weighted graph, $G = (V,E,J)$, on $n$ nodes, where node $v_i$ has node weight $J_{ii}$ and edge $e_{ij}$  has edge weight $J_{ij}$. Using QAOA to solve the above problem will require, at minimum, $n$ qubits. Furthermore, every iteration of QAOA will require $n$ single qubit gates and $|E|$ two qubit gates \cite{herrman2021lower}. 

Classical optimization techniques can be very effective at solving QUBO instances. If a quantum advantage can be seen in QUBO, it will be on large instances (both in terms of $n$ and $|E|$). Unfortunately, near-term quantum computers do not have the hardware needed to solve these large problems (both in terms of the number of qubits and the fidelity needed to perform $p|E|$-many two qubit gate operations). With that in mind, we attempt to solve a large QUBO by instead solving many smaller QUBOs, each of these requiring far fewer qubits and fewer gates.

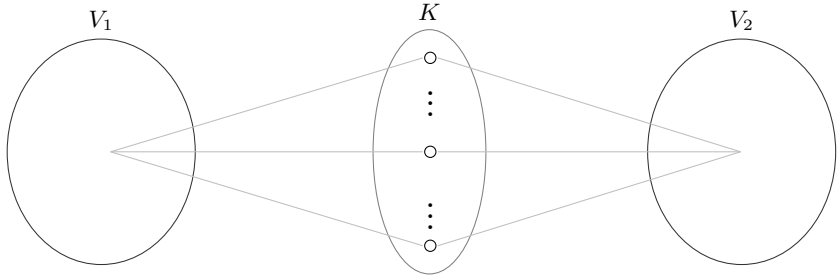
\begin{figure}
    \centering
    \begin{tikzpicture}[scale=0.5,node distance = 5mm, E/.style = {ellipse, draw=#1, minimum height=3.0cm, minimum width=2.5cm},
    E/.default=black!80!white]
    \node (e1) [E,
                label=above: $V_1$] {};
    \node (e3) [E, right=0.0cm and 6cm of e1, label=above: $V_2$] {};
    \node (e2) [E=black!50!white, right=0.0cm and 2.35cm of e1, label=above: $K$, minimum width=1.5cm, minimum height=3.25cm] {};
    \draw (8.75,2.5) circle (0.15cm);
    \draw (8.75,1.5) node [black] {\textbf{\vdots}};
    \draw (8.75,0) circle (0.15cm);
    \draw (8.75,-1.5) node [black] {\textbf{\vdots}};
    \draw (8.75,-2.5) circle (0.15cm);
    \draw[lightgray] (0.25,0) -- (8.55,0);
    \draw[lightgray] (8.95,0) -- (17,0);
    
    \draw[lightgray] (0.25,0) -- (8.55,2.5);
    \draw[lightgray] (8.95,2.5) -- (17,0);
    
    \draw[lightgray] (0.25,0) -- (8.55,-2.5);
    \draw[lightgray] (8.95,-2.5) -- (17,0);
    \end{tikzpicture}
    
    \caption{A graph with cut set $K$ of size $k$ that splits the graph into two components, $V_1$ and $V_2$. }\label{fig:cutsets}
\end{figure}

We can create a {\em restriction} of~\eqref{eq:qubo} by fixing a collection of variables to some values. Let $F_1$ denote the collection of variables fixed to one and $F_0$ the set of variables fixed to zero. The restriction of~\eqref{eq:qubo} with respect to $(F_0, F_1)$ is the following 
\begin{align}
C_{max F_0,F_1} = \max_{z\in \{0,1\}^n|_{F_k}} \sum_i \sum_{j \neq i} J_{ij}z_i z_j  + \sum_i J_{ii} z_i. \label{eq:restrict_qubo}
\end{align}

\noindent Note that we use ${z\in \{0,1\}^n|_{F_k}}$ to denote values for $x$ such that $z_i = 1$ for all $i$ in $F_0$ and $z_i = 1$ for all $i$ in $F_1$.  Note that for any $(F_0, F_1)$, we have that $C_{max F_0, F_1} \leq C_{max}$ 

Let $K$ be a cut set of $G$ such that its removal leaves two components that have vertex sets $V_1$ and $V_2$ (note that $V_1 \cup V_2 \cup K = V$). For any restriction $(F_0, F_1)$ where $F_0 \cup F_1 = K$, the restricted problem can be written as:
\begin{align}
C_{max F_0, F_1} = \max_{z\in \{0,1\}^n|_{F_k}}   \sum_{l = 1}^2 \big( \sum_{i \in V_l} \sum_{j \in V_l \setminus \{i\}}   J_{ij}z_i z_j  +  \sum_{i \in V_l} \overline{J}_{ii} z_i \big) +    c, \label{eq:decomp}
\end{align}

\noindent where
$\overline{J_{ii}}  = J_{ii} + \sum_{j \in F_1}  J_{ij}$ and 
$c$ is the constant term $\sum_{i \in F_1}\sum_{j \in F_1 \setminus\{i\}} J_{ij}   +  \sum_{i \in F_1} J_{ii}$. When applying a restriction $(F_0, F_1)$. Note that for this restriction, the variables representing vertices in $V_1$ are independent of those in $V_2$, meaning that:

$$C_{max F_0, F_1} = C_{max F_0, F_1}^1 + C_{max F_0, F_1}^2 + c, $$

\noindent 
where
\begin{align}
C_{max F_0, F_1}^1 = \max_z &      \sum_{i \in V_1} \sum_{j \in V_1\setminus\{i\}}   J_{ij}z_i z_j  +  \sum_{i \in V_1} \overline{J}_{ii} z_i \label{eq:sub1}\\
C_{max F_0, F_1}^2 = \max_z &      \sum_{i \in V_2} \sum_{j \in V_2\setminus\{i\}}   J_{ij}z_i z_j  +  \sum_{i \in V_2} \overline{J}_{ii} z_i. \label{eq:sub2}
\end{align}

Thus, any restriction formed by fixing variables in a cut set can be decomposed into two smaller, independent problems. If one could find the ``correct'' values of the variables  in $K$, those that correspond with the optimal solution, then solving the two subproblems would be equivalent to solving the original problem. In practice, it is unlikely that one would be able to practically (a) find a cut set that decomposes a hard problem into two easy problems and (b) find the correct restriction for that cut set. What is often the case is that one subproblem is very large and difficult while the other subproblem is small and easy. Without loss of generality, we can assume that solving for $C_{max F_0, F_1}^1$ is harder than solving for $C_{max F_0, F_1}^2.$ In order to create a new problem instance that only include the $V_1$ and $K$ variables, and whose optimal solution value is identical to the original QUBO, we compute new coefficients $\hat{J_{ij}}$ and $\hat{J_{ii}}$ for $i,\ j\ \in V_1 \cup K$ and constant $\hat{c}$ such that

\begin{align} C_{max} = \max_{z \in V_1 \cup K} \sum_{i\in V_1 \cup K} \sum_{j\in V_1 \cup K\setminus\{i\}} \hat{J_{ij}}z_i z_j  + \sum_{i\in V_1 \cup K} \hat{J_{ii}} z_i + \hat{c}. 
\label{eq:goal}
\end{align}

\noindent Note that the right hand side contains just variables in $V_1$ and $ K $ while the original problem includes all vertices. We can interpret $\hat{J}$ as the adjacency matrix of a new, smaller, weighted graph. 

We can attempt to find such values by the following. Let $s \in \{0,1\}^{|K|}$ be a bit string representing the values of variables in $K$, and let $C_{max_s}$ be the optimal solution to~\eqref{eq:sub2} and $c$ the constant term from~\ref{eq:decomp} where the variables in $K$ are fixed to the value given in $s$. We attempt to find  $\hat{J_{ij}}$, $\hat{J_{ii}}$, and $\hat{c}$ that satisfy the following system of equations.

\begin{align}\label{eq:soq}
    \sum_{ \{i, j | s[i] = s[j] = 1\} } \hat{J_{ij}} +  \sum_{ \{i | s[i] = 1\} } \hat{J_{ii}} + \hat{c} = C_{max_s}  +c &\  \forall s \in \{0,1\}^{|K|}.
\end{align}

If the above system of equations is not feasible, then we would like to determine values of $\hat{J}$, $\hat{J_{ii}}$, and $\hat{c}$ such that the left hand side of~\eqref{eq:goal} is approximately equal to the right hand side. We can attempt to find such values by introducing error terms to~\eqref{eq:soq} and instead finding solutions to the following system of equations: 

\begin{align} \label{eq:soq_approx}
    \sum_{ \{i, j | s[i] =s[j] = 1\} } \hat{J_{ij}} +  \sum_{ \{i | s[i] = 1\} } \hat{J_{ii}}  + \hat{c} +  e_s= z_s &\  \forall s \in \{0,1\}^{|K|}, 
\end{align}

\noindent where $e_s$ indicates the error associated with the partial solution $s$. 
Note that~\eqref{eq:soq_approx} is always feasible (as $e_s = C_{max_s}$ and all other variables equal to zero is a solution). With that in mind, we want to find $\hat{J}$, $\hat{J_{ii}}$, and $\hat{c}$ that minimize the error. Note that one can use any metric to measure the error. Throughout this paper we will minimize the sum of all the error terms with the additional constraint that each $e_s \geq 0$. The lower bound of no error can be achieved for graphs that have minimum cut sets of size at most three.

\begin{theorem}\label{thm:reweight}
    Let  $G = (V,E,J)$ be a weighted graph with cut set  $K$,  $ |K| \leq 3$. Let $V_1$ and $V_2$ be the vertices in the resulting connected components caused by cut $K$.  The proposed approach generates  a weighted graph $G'(V_1 \cup K, E',w')$ such that $C_{max}$ is the value for the optimal bit string for a MaxCut solution to $G$ if and only if $C_(z)_{G'}$ (the projection of $C(z)$ onto the vertices of $G'$ by deleting entries representing vertices in $V_2$) is an optimal solution to the MaxCut problem on $G'$. 
\end{theorem}

\begin{proof}
It is easy to verify that the systems of equations in \eqref{eq:soq} are full rank and invertible for $|K|\leq 3$, since there are three linearly independent equations with three unknowns, giving a unique solution to the $\hat{J}$ values.
\end{proof}


\noindent For $K > 3$, it is possible that $e_s > 0$. 

\subsection{Algorithm for determining $K$, $\hat{J}$ and $\hat{c}$}

The formal description of the algorithm is presented in Algorithms~\ref{alg:DecompAlg}-~\ref{alg:weight}. Note that at least one vertex is removed from the graph at each iteration and that the algorithm terminates when either the smallest cut set is greater than some predetermined value or if the resulting graph contains just one vertex. As a result, we know the decomposition algorithm will require at most $|V|$-many iterations. In addition, finding minimum cut sets can be found in polynomial time~\cite{ford1956maximal}. However, solving the MaxCut problem for each of the subproblems is exponential in the size of $V_2$. Note, however, that we can use heuristics to approximate the objective values for these problem instances. 

The input of the Algorithm~\ref{alg:DecompAlg} is a graph (either weighted or unweighted) that contains some cut set $K$ that splits the graph into two components and some $M$, where the algorithm terminates if there are no cut sets of size smaller than $M$. Its output is some weighted graph with strictly fewer vertices than the original graph that has an objective value equal to or bounded above by the optimal objective value of the original graph. MaxCut on the resulting decomposed graph can be solved using fewer qubits than the original problem on a quantum device.
To achieve this end, there is a subroutine Algorithm~\ref{alg:weight} that removes a set of vertices $V_2$. By removing these set of vertices, all possible edges whose endpoints are in $K$ are reweighted and $\hat{c}$ is obtained such that the optimal MaxCut solution of the decomposed graph alongside $\hat{c}$ is bounded above by the optimal solution of the input graph.

\begin{algorithm}[H]
\caption{Decomposition Algorithm}
\begin{algorithmic}
    \Input{$G^{(0)}=(V^{(0)},E^{(0)},J^{(0)})$, M}
    \Function{Decomp}{$G^{(0)}$}
      \State initialize $i \leftarrow 0$ and $c \leftarrow 0$ and $G^{(i)} \leftarrow G^{(0)}$
        \State $K^{(i)} \leftarrow$ minimum cut set of $G^{(i)}$, $V^{(i)}_1$, $V^{(i)}_2 \leftarrow G \backslash K$ \Comment{$|V^{(i)}_1| \geq |V^{(i)}_2|$}
        \State $J^{(i)} \leftarrow$ adjacency matrix of $G$
      \While{$|K^{(i)}|  < M $} \Comment{$M$ is a predetermined parameter}
        \Function{Reweight}{$K^{(i)} \cup V^{(i)}_2$} 
            \State \Return $\hat{J}$, $\hat{c}$
        \EndFunction
        \For{$u \neq v \in K\times K$}
            \If{$uv \in V^{(i)}_1 \cup K^{(i)}$}
                \State $J^{(i+1)}_{uv} \leftarrow J_{uv}^{(i)}$
                \ElsIf{$uv \in K^{(i)} \times K^{(i)}$}
                \State  $J_{uv}^{(i+1)} \leftarrow \hat{J}_{uv}$
            \EndIf
        \EndFor
        \State $V^{i+1} = \{V_{1}^{(i)} \cup K^{(i)} \}$
        \State $E^{(i+1)} = E(V^{(i)}_{1}) \cup E(K^{(i)} \times K^{(i)})$
        \State $\mathcal{G}^{(i)} = (V^{(i+1)}, E^{(i+1)}, J^{(i+1)})$
        \State $G^{(i+1)} \leftarrow \mathcal{G}^{i}$
        \State $\hat{c^{(i)}} \leftarrow \hat{c}$
        \State increment $i$
      \EndWhile
      \State \Return $\mathcal{G}$, $\hat{c}$
    \EndFunction
    \Output{$\mathcal{G}$, $\hat{c}$}
\end{algorithmic}\label{alg:DecompAlg}
\end{algorithm}

\begin{algorithm}[H]
  \begin{algorithmic}[1]
    \Input{$G^{(i)}$, $K^{(i)}$, $J^{(i)}$, $J^{(i)}$, $V^{(i)}_1$ $V^{(i)}_2$, $i$}
    \Function{Reweight}{G}
    \State $H  \leftarrow$ induced graph generated from $K \cup V_2$
    \State Generate an array $S$ that contains all possible $s \in \{0,1\}^{|K|}$ 
    \State initialize $\hat{c} \leftarrow 0$
    \State initialize $\vec{b} \leftarrow \vec{0}$
    \For{$ind \in range(S)$}
        \State Fix the qubits in $k_l \in K$ to either 0 or 1 with respect to the $l^{\text{th}}$ entry in $S[ind]$
        \State $C_s \leftarrow$MaxCut objective value with fixed qubits in set $K$ 
 
    \State $\vec{b}[ind] = C_s$ 
    \EndFor
   
    \State Solve for~\eqref{eq:soq_approx} to find $\hat{J}$
     \State $\hat{c} \pm \hat{c}^{(i)}$
      \State \Return {$\hat{J}$, $\hat{c}$}
    \EndFunction
    \Output{$\hat{J}$, $\hat{c}$}
  \end{algorithmic}
  \caption{Reweight Algorithm}
  \label{alg:weight}
\end{algorithm}

    
    
    

\subsection{Example: Decomposition approach }\label{subsec:graphdecompexample}

\begin{figure}
    \centering
    \begin{subfigure}{0.3\textwidth}
            \begin{tikzpicture}[scale = 0.6]
            \Vertex [IdAsLabel, y=1] {1} \Vertex[IdAsLabel,x=3,y=3]{2} \Vertex[IdAsLabel, x=3, y =1]{3} \Vertex[IdAsLabel, x=3,y=-1]{4}
            \Vertex[IdAsLabel, x=6, y = 3]{5} \Vertex[IdAsLabel, x=6, y = -1]{6}
            
            \Edge(1)(2)
            \Edge(1)(3)
            \Edge(1)(4)
            \Edge(4)(6)
            \Edge(4)(5)
            \Edge(3)(6)
            \Edge(3)(5)
            \Edge(2)(6)
            \Edge(2)(5)
            \end{tikzpicture}
    \end{subfigure}%
    \begin{subfigure}{0.3\textwidth}
            \begin{tikzpicture}[scale = 0.6]
            \Vertex [IdAsLabel, y=1] {1} 
            \Vertex[IdAsLabel, x=6, y = 3]{5} \Vertex[IdAsLabel, x=6, y = -1]{6}
            
            \end{tikzpicture}
    \end{subfigure}
    \begin{subfigure}{0.3\textwidth}
             \begin{tikzpicture}[scale = 0.6]
            \Vertex [IdAsLabel, y=1] {1} \Vertex[IdAsLabel,x=3,y=3]{2} \Vertex[IdAsLabel, x=3, y =1]{3} \Vertex[IdAsLabel, x=3,y=-1]{4}
            
            \Edge(1)(2)
            \Edge(1)(3)
            \Edge(1)(4)
            \end{tikzpicture}
    \end{subfigure}
    \caption{Left: Graph $G$. Middle: $G$ is split into two components from the removal of $K$. Right:  $V_2 \cup K$}
    \label{fig:embedding}
\end{figure}
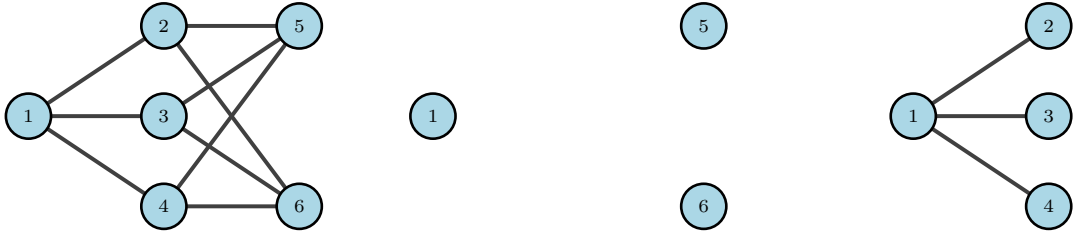

We now give an example of  the decomposition approach applied to a graph, $G = (V,E,J)$, pictured in Fig.~\ref{fig:embedding}. $G$ has a minimum cut set $K = \{2,3,4\}$ whose removal splits the vertices of $G$ into two nonempty sets: $V_1=\{5,6\}$ and $V_2=\{1\}$, such that $|V_1| \geq |V_2|$. Algorithm~\ref{alg:DecompAlg} finds this $K$ and then calls Algorithm~\ref{alg:weight} as a subroutine to remove all vertices in $V_2$, add edges in $K$, and re-weight so the new graph has the same objective value as the original. Algorithm~\ref{alg:DecompAlg} will iterate this process until it reaches a termination criteria.

Next (Algorithm~\ref{alg:weight}), 
we solve the subproblem on the subgraph generated by the vertices in $V_2$ and $K$ for all of the $2^{|K|}$ possible values of vertices in $K$. This is shown in Table~\ref{table:example3sp1solutions}. Using these values, we set up the system of equations as shown in~\eqref{eq:soq} to find the weights of the new edges in $K$. Note that since $|K|$ is three, we know there is an exact  solution to the reweighting problem, so we can solve the system of equations~\eqref{eq:soq} instead of the linear program~\eqref{eq:soq_approx}. 

\begin{align}  
 & \hat{c} = 3 & (s = 000)\\
 & \hat{J}_{24} + \hat{J}_{34} + \hat{c}= 2 & (s = 001)\\
 & \hat{J}_{23} + \hat{J}_{34} + \hat{c} = 2 & (s = 010)\\
  & \hat{J}_{23} + \hat{J}_{24} + \hat{c} = 2 & (s = 100)
  \end{align}

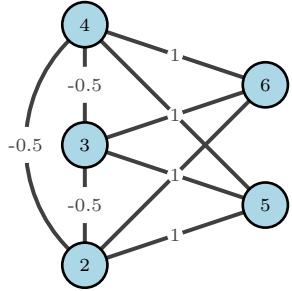
\begin{figure}
    \centering
    \begin{tikzpicture}[scale = 0.8]
    \Vertex[IdAsLabel,x=3,y=-1]{2} \Vertex[IdAsLabel, x=3, y =1]{3} \Vertex[IdAsLabel, x=3,y=3]{4}
    \Vertex[IdAsLabel, x=6, y = 0]{5} \Vertex[IdAsLabel, x=6, y =2]{6}
    \Edge[label=1](4)(6)
    \Edge[label=1](4)(5)
    \Edge[label=1](3)(6)
    \Edge[label=1](3)(5)
    \Edge[label=1](2)(6)
    \Edge[label=1](2)(5)
    \Edge[bend=-45, label = -0.5](4)(2)
    \Edge[label = -0.5](4)(3)
    \Edge[label = -0.5](3)(2)
    \end{tikzpicture}
    \caption{Graph $\mathcal{G}$ for 3-Cut Example }
    \label{fig:exampledecomp}
\end{figure}

\noindent Solving the above system of equations gives us $\{ \hat{J}_{23} = -0.5, \hat{J}_{24} = -0.5, \hat{J}_{34} = -0.5, \hat{c} = 3\}$. We use this to modify the original graph by removing node $A$ and all incident edges and adding an edges $23$, $24$, and $34$ each with weight -0.5. The constant $\hat{c}=3$ is added to the objective function. The new graph is shown in Figure~\ref{fig:exampledecomp}. Note that the optimal solution to the new graph ($x_5 = x_6 = 1$ and $x_2 = x_3 = x_4 =0$) has objective value of 6. That, plus the constant term of 3 gives 9, which is the optimal solution to the original graph.

\begin{table}
    \centering
    \begin{tabular}{ccccc|c}
        & \multicolumn{3}{c}{\makecell{Partial\\ Assignments}} & & \multicolumn{1}{c}{} \\%
     \arrayrulecolor{Gainsboro} \cmidrule[\heavyrulewidth](l){2-4}\cmidrule[\heavyrulewidth]{6-6} \arrayrulecolor{black}
        & \makecell{\; $V_2$ \;}&\makecell{\; $V_3$ \;} & \makecell{\; $V_4$ \;} & & \makecell{Subproblem\\Solution} \\
    \midrule
        & 0 & 0 & 0 &  & 3.0 \\
        & 0 & 0 & 1 &  & 2.0 \\
        & 0 & 1 & 0 &  & 2.0 \\
        & 1 & 0 & 0 &  & 2.0 \\
        \bottomrule
    \end{tabular}
        \begin{tabular}{ccccc|c}
        & \multicolumn{3}{c}{\makecell{Partial\\ Assignments}} & & \multicolumn{1}{c}{} \\%
     \arrayrulecolor{Gainsboro} \cmidrule[\heavyrulewidth](l){2-4}\cmidrule[\heavyrulewidth]{6-6} \arrayrulecolor{black}
        & \makecell{\; $V_2$ \;}&\makecell{\; $V_3$ \;} & \makecell{\; $V_4$ \;} & & \makecell{Subproblem\\Solution} \\
    \midrule
        & 0 & 1 & 1 &  & 2.0 \\
        & 1 & 1 & 0 &  & 2.0 \\
        & 1 & 0 & 1 &  & 2.0 \\
        & 1 & 1 & 1 &  & 3.0 \\
        \bottomrule
    \end{tabular}
    \caption{All MaxCut partial assignment subproblem solutions for 3-cut example.}
    \label{table:example3sp1solutions}
\end{table}

\section{Implementation and Results}\label{sec:qaoadecomp}


We apply the above decomposition algorithm to twenty-five 3-regular 100-vertex MaxCut instances. 

\subsection{Algorithm implementation details}

In order to implement Algorithm~\ref{alg:DecompAlg}, we use max-flow implementation~\cite{networkX,esfahanian2013connectivity} to find a minimum cut set, $K$, of each input graph. The two sets into which the graph is partitioned when the cut set is removed are called $V_1$ and $V_2$, where without loss of generality, $|V_1| \geq |V_2|$. Algorithm~\ref{alg:DecompAlg} then calls on Algorithm~\ref{alg:weight}. In Algorithm~\ref{alg:weight}, each vertex in $K$ is assigned a value $0$ and $1$, and the MaxCut objective for $V_2 \cup K$ is calculated. We test two different approaches to solving these subproblems. The purely classical approach uses Gurobi\cite{gurobi}, a state-of-the art commercial software. We also test using QAOA to find solutions 
 We note that Gurobi returns an exact optimal solution while QAOA returns a heuristic solution based on the expected cost value, as we now describe. 

The expected value of general Ising problems was derived by Ozaeta, van Dam, and McMahon \cite{Ozaeta_2022}. These formulas, found in Appendix~\ref{app:formula}, were used to calculate the expected value of the subproblem when solved with QAOA \cite{Ozaeta_2022}. The optimal parameters were selected by maximizing the sum of the expected value terms for each problem using the Broyden-Fletcher-Goldfarb-Shanno (BFGS) algorithm \cite{wright1999numerical}. As BFGS terminates in a local minimum, we repeated each optimization 100 times using different random starting parameters.

Once the subproblems are solved, we use Gurobi to solve equation Eq.~\eqref{eq:goal}, and thus calculate the changes of weights on edges incident to $K$. Note that after edge-weights are updated, the graph is no longer guaranteed to remain 3-regular. The process continues until a termination criteria is met. For our test runs, Algorithm~\ref{alg:DecompAlg} terminated when either $|N|\leq 2$ or $|K| > 7$.  

\begin{figure}
    \centering
    \begin{subfigure}{0.5\textwidth}
    \includegraphics[width=0.9\linewidth]{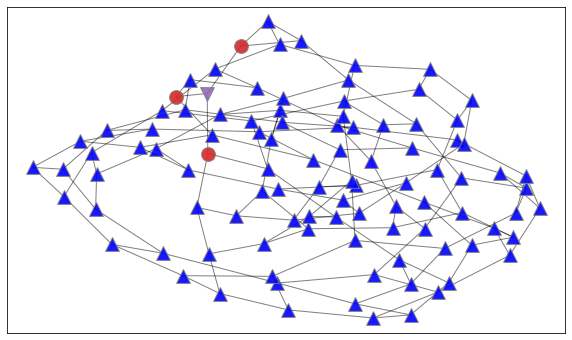}
    \end{subfigure}%
    \begin{subfigure}{0.5\textwidth}
    \includegraphics[width=0.9\linewidth]{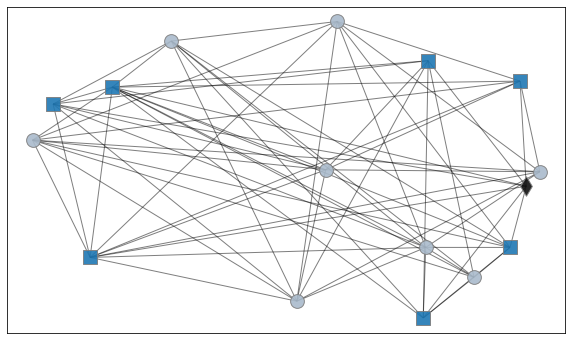}
    \end{subfigure}
    \caption{An example of one of the 100-vertex 3-regular graph solved using QAOA and then used in the decomposition algorithm (left). Marked in the red (circle): $K^{(0)}$, in blue (upward-facing triangle): $V^{(0)}_1$, and in
    purple (down-facing triangle): $V^{(0)}_2$. The final iteration of the decomposition before the termination criteria is met (right). Marked in the light blue (square): $K^{(84)}$, in gray (circle): $V^{(84)}_1$, and in
    black (rhombus): $V^{(84)}_2$. }
    \label{fig:decompositionIterations}
\end{figure}
Fig.~\ref{fig:decompositionIterations} is an example of one of the graphs solved using Algorithm~\ref{alg:DecompAlg}. The algorithm was applied until the termination criteria was met, which can be easily verified since each vertex has degree at least seven, so at least seven vertices must be removed from the graph to isolate at least one vertex. After 85 iterations, the 100 vertex graph decomposed into a 15 vertex graph which requires substantially fewer qubits to solve.

\subsection{Results}
\begin{figure}
    \centering
    \includegraphics[scale=0.55]{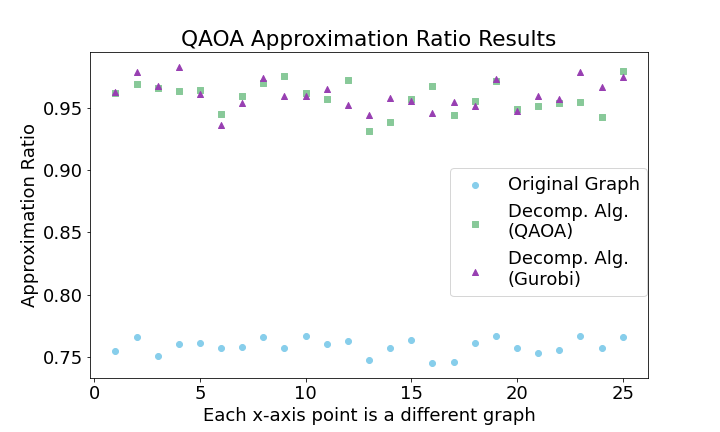}
    \caption{Comparing approximation ratios for the twenty-five different 100-vertex 3-regular random graphs. }
    \label{fig:50vertex3-RegApprox}
\end{figure}

The primary metric of QAOA success is the approximation ratio, which is the expected value of the solution output by QAOA divided by the optimal solution value.  Fig.~\ref{fig:50vertex3-RegApprox} gives the approximation ratio for 25 different 100-vertex MaxCut problem instances using  QAOA on the original graphs compared to using QAOA on the decomposed graphs. We tested using both Gurobi and QAOA to solve the subproblems. On average, the approximation ratio when the original problems are solved with QAOA is $0.758452$. When the same problems are solved using decomposition to augment QAOA, the average approximation ratio is $0.961040$ for the Gurobi decomposed instances and $0.95865$ for the QAOA decomposed instances. 
In general, QAOA augmented with either the QAOA or Gurobi decomposition methods outputs solutions much closer to the optimal MaxCut solution than the QAOA solution without decomposition. Interestingly, while using Gurobi tends to produce the best approximation ratios, there is not a considerable difference between Gurobi and QAOA decomposition.  



\begin{figure}
    \centering
    \includegraphics[width=0.9\linewidth]{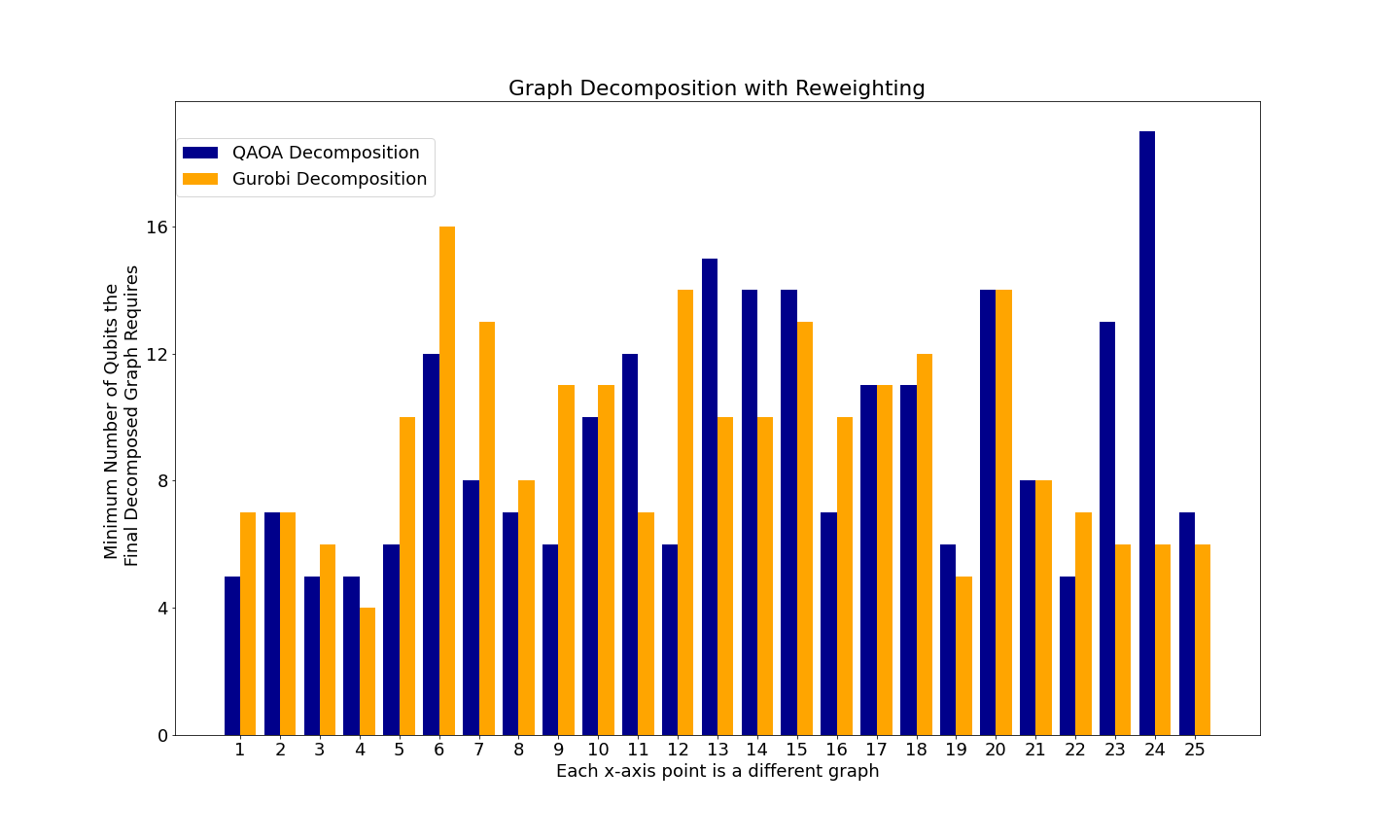}
    \caption{The minimum number of qubits needed for each of the final decomposed graphs using QAOA and Gurobi for the reweighting. }
    \label{fig:numQub}
\end{figure}

Using decomposition techniques with QAOA not only increases the average approximation ratio for these graphs, but it also decreases the number of qubits needed to solve the problem. Each problem originally required 100 qubits to solve, since each graph has 100 vertices. On average, the decomposed graphs can be solved with QAOA using 9.32 qubits to solve the QAOA reduced problems and 9.28 qubits to solve the Gurobi reduced problems. This is a 90.68\%(QAOA) decrease on the number of qubits in the former cases and a 90.72\%(Gurobi) decrease in the latter cases, on average. Fig.~\ref{fig:numQub} shows how many qubits QAOA requires to solve MaxCut for each of the decomposed problem instances.







\begin{figure}
    \centering
    \includegraphics[scale=0.55]{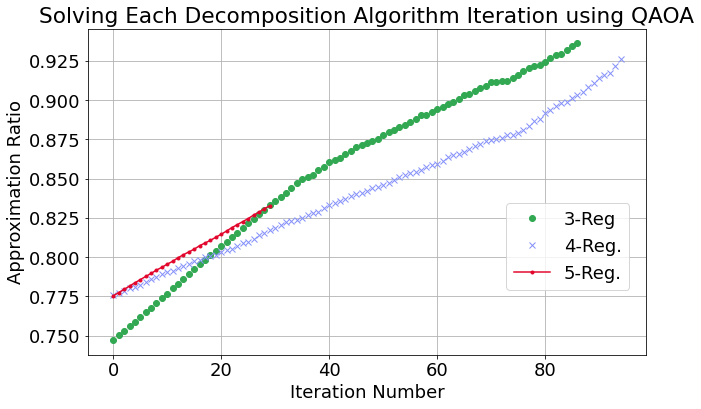}
    \caption{Showing QAOA applied for each iteration of the decomposition algorithm.}
    \label{fig:qaoadecomp_iterations}
\end{figure}

In the first experiment, we ran Algorithm~\ref{alg:DecompAlg} until the termination criteria that $|K| > 7$ was met. The number of iterations it took to meet the criteria differed for each graph, so the impact of each decomposition iteration was not captured. In order to determine how the number of iterations affects the approximation ratio, we ran the decomposition algorithm on a single 3-regular graph in the above study, as well as a single 100-vertex, 4-regular graph, and a single 100-vertex, 5-regular graph. After each iteration of Algorithm~\ref{alg:DecompAlg}, we solved the MaxCut problem on the reduced graph with QAOA and calculated the approximation ratio. This process was iterated until the termination criteria was met. The plot of approximation ratio as a function of decomposition iterations is found in Fig.~\ref{fig:qaoadecomp_iterations}. The number of decomposition iterations for the 3-regular graph was eighty-five. The 4-regular graph terminated after eighty-seven iterations, and the 5-regular graph terminated after twenty-seven iterations. As seen in the plot, each line is monotonically increasing, which is to be expected since the number of vertices in the graph decreases with each iteration. 

For the three-, four-, and 5-regular graphs, the termination criteria $|K| > 7$ was met after eighty-six, ninety-five, and thirty iterations, respectively. The 3-regular graph reaches an approximation ratio of approximately 0.93601 after the eighty-six iterations, where the original graph had an approximation ratio of 0.74724. The 4-regular graph originally had an approximation ratio of 0.77559. After ninety-five iteration of the decomposition, it had an approximation ratio of 0.92628. Originally, the 5-regular graph has an approximation ratio of 0.77533. After thirty iterations, the decomposed graph had an approximation ratio of 0.83267. 
Note that the potential reduction obtained by the algorithm can be bounded below for $k$-regular graphs. For any $v \in V$ the nodes adjacent to $v$ form a cut set of size $k$. Using this as our cut will give $V_2 = \{v\}$. The decomposition algorithm will remove vertex $v$ and in the worst case increase the degree of $k$-many neighbors by one, leaving other nodes unchanged. In aggregate, the degrees of all but $k+1$ nodes are unchanged. As a result, there will always be at least one degree-$k$ node until $\lceil \frac{n}{k+1}\rceil$-many iterations have been performed where one vertex is removed. As a result, there will be at most $\frac{k}{k+1}n$ vertices remaining. 

Finally we examine the performance of our approach in quantum computations using the Quantinuum trapped-ion quantum computer H1-1.  We selected ten 100-vertex graphs that were each reduced to between 12 and 16 qubits following our decomposition.  We ran $p=1$ QAOA circuits for each of these graphs using the software \texttt{tket}\cite{sivarajah2020t} to submit jobs to H1-1.  We defined circuits in terms of CNOT, rotation, and Hadamard gates following a standard approach \cite{lotshaw2022scaling}, then optimized these circuits and compiled them to the native gate set of H1-1 using \texttt{tket}.  We took 500 shots per instance to sample varying solutions to these combinatorial problems. Table~\ref{tab:results} in Appendix~\ref{sec:appendixresults} compares the approximation ratio from the decomposition simulations to the H1-1 test runs.

\begin{figure}
    \centering
    \includegraphics[width=13cm,height=12cm,keepaspectratio]{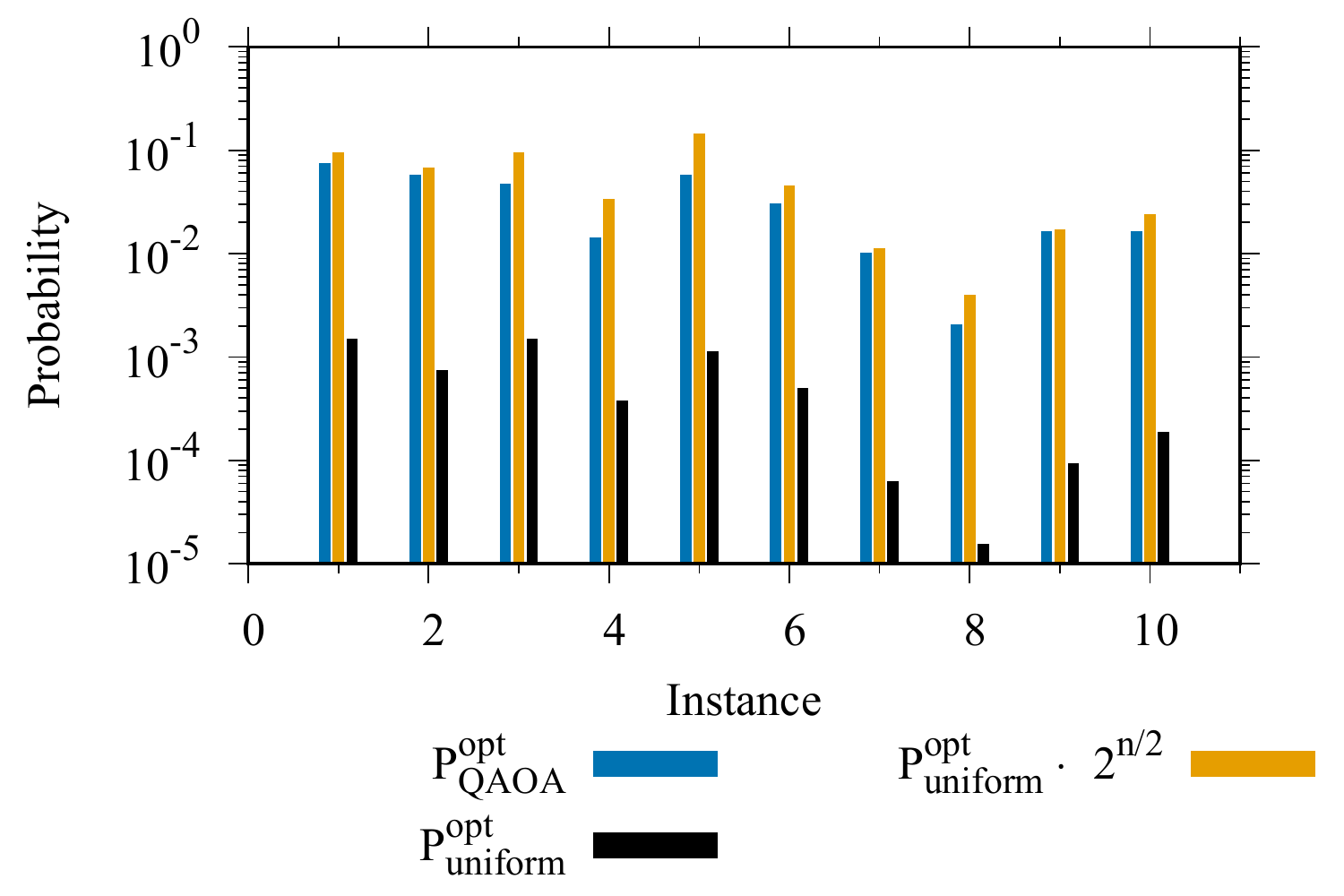}
    \caption{Optimal solution probabilities for ten reduced 100-vertex problem instances. Probabilities $P_\mathrm{QAOA}^\mathrm{opt}$ are sampled from the quantum computer H1-1 running $p=1$ QAOA circuits with 500 shots, probabilities $P_\mathrm{uniform}^\mathrm{opt}$ are from the uniform distribution on the reduced problems, and $P_\mathrm{uniform}^\mathrm{opt}\times 2^{n/2}$ are probabilities when each individual optimal solution probability is enhanced by a factor $2^{n/2}$. }
    \label{H1-1}
\end{figure}
Figure \ref{H1-1} shows the optimal solution probabilities $P_\mathrm{QAOA}^\mathrm{opt}$ we observed in these quantum computations (blue).  In each problem the quantum computer succeeded in observing the optimal solution within 500 shots.  For comparison, we plot the probabilities $P_\mathrm{uniform}^\mathrm{opt}$ that are expected from the uniform distribution (orange) on the same reduced graphs, corresponding to the uniform superposition in the QAOA initial state. Here $P_\mathrm{uniform}^\mathrm{opt} = N^\mathrm{opt}/2^n$, where $N^\mathrm{opt}$ is the number of optimal solutions in the reduced problem and $n$ is the corresponding number of qubits.  In comparison, the $p=1$ QAOA circuits succeed in significantly enhancing the optimal solution probabilities.  It is important to note that the equivalent non-reduced problems have optimal solution probabilities $\sim 2^{-100} \approx 10^{-10}$ in their uniform distributions, since they have 100 qubits.  Reducing these problems considerably increases the likelihood of identifying optimal solutions in the uniform distribution, and $p=1$ QAOA gives a further significant increase to the probabilities as observed in the quantum computations on H1-1. 

For a final comparison, we computed probabilities $P_\mathrm{uniform}^\mathrm{opt}\times 2^{n/2}$ (orange) in Fig.~\ref{H1-1}.  Probability enhancements $\sim 2^{n/2}$ were observed numerically for a variety of $p=1$ QAOA instances in previous work \cite{DiezValle2023pseudoBoltzmann} and we find the QAOA probabilities from H1-1 are close to the $P_\mathrm{uniform}^\mathrm{opt}\times 2^{n/2}$, consonant with this previous work.  Overall, we conclude from these quantum computations that our graph reductions are a powerful technique for increasing the optimal solution probability with QAOA, and that a current quantum computer is capable of consistently finding optimal solutions for 100-qubit problems after using our techniques to define equivalent reduced problems with fewer qubits.

\section{Discussion}\label{sec:discussion}
In this work, we develop a decomposition algorithm that is able to significantly reduce the quantum hardware required as well as increase the performance of quantum algorithms.  We decompose twenty-five 3-regular, 100-vertex MaxCut instances using Gurobi and QAOA, and then solve these decomposed problems using one iteration of QAOA. We compare the approximation ratios of these methods to the QAOA approximation ratio when solving the original instances. The decomposed approximation ratios are on average about 26\%  higher than the approximation ratio obtained when solving the original problem with QAOA for Gurobi and QAOA decompositions, respectively. QAOA requires 100 qubits to solve these problems on fully-connected hardware, since there are 100 vertices in each instance. When using QAOA to solve the reduced problems on the same hardware, the algorithm requires, on average 9 qubits for both Gurobi and QAOA reduced which represents a 90\% decrease.  This qubit reduction is especially significant for current quantum devices, which have a limited number of qubits and may have architecture restrictions, and the increase in approximation ratio indicates that fewer iterations of the algorithm will be needed to find a high-quality solution. We confirmed this by executing $p=1$ QAOA computations on a trapped-ion quantum computer, which succeeded in finding optimal solutions with significant probability for ten 100-qubit problem instances that we decomposed into equivalent problems with 12-16 qubits. 

While these results are significant, they were obtained only on a small collection of graphs that have similar structure. Future work includes using decomposition techniques on a wider variety of graphs and analytically determining approximation ratio bounds for solving decomposed problems. Furthermore, it would be of interest to determine how the decomposition techniques can be used in conjunction with QAOA variations such as ma-QAOA \cite{herrman2022multi}. Additional avenues of future work would determine if similar decomposition techniques can be applied to other combinatorial optimization problems, and if solving the reduced problems results in similar approximation ratio increases while requiring fewer qubits. Finally, it would be interesting to compare QAOA$^2$ \cite{zhou2023qaoa} and ML-QLS \cite{ushijima2021multilevel} performance to this decomposition method the same set of test graphs to determine if one approach has an advantage over the other, since previous QAOA$^2$ work studied only smaller MaxCut instances and ML-QLS work studied the graph partitioning problem and and modularity maximization problem.


\acknowledgments
This work was supported by DARPA ONISQ program under award W911NF-20-2-0051. G.S.\ acknowledges support by the National Science Foundation under award DGE-2152168 and  the Army Research Office under award W911NF-19-1-0397. This research used resources of the Oak Ridge Leadership Computing Facility, which is a DOE Office of Science User Facility supported under Contract DE-AC05-00OR22725.

\renewcommand\refname{References Cited}
\bibliography{Main}
\bibliographystyle{IEEEtran}

\newpage

\appendix
\section{\\Closed form for the expected value of the $C_i$ and $C_{ij}$}\label{app:formula}

These are the equations we use to calculate the expected energy of the objective value, which is an important step in Alg.~\ref{alg:weight} as each possible restriction problem will need to be solved to obtain $c$ and some $\vec{b}$ that contains the new $\hat{J}$. 

The first equation is the Ising formula for the calculating the vertex terms in the graph. The second equation is used in calculating the edge terms in the graph\cite{ozaeta2022expectation}.

\begin{equation*}
  \langle C_{i} \rangle =  J_{ii} \sin(2 \beta) \sin(2 \gamma )  \prod_{(ik) \in E} \cos(2 \gamma J_{ik}) 
\end{equation*}
\noindent and

\begin{align*}
     \langle C_{ij} \rangle &=  \frac{J_{ij}\sin(4\beta)}{2} \sin(2 \gamma J_{ij}) \big[\ \cos(2\gamma J_{ii}) \prod_{\substack{(i,j) \in E \\ k\neq j} } \cos(2\gamma J_{ik}) + \cos(2\gamma J_{jj}) \prod_{\substack{(j,k) \in E \\ k\neq i} } \cos(2\gamma J_{jk}) \big]\\\
      &  - \frac{J_{ij}}{2} (\sin(2 \beta))^{2} \prod_{\substack{(ik) \in E \\ (jk) \notin E} } \cos(2 \gamma J_{ik}) \prod_{\substack{(jk) \in E \\ (ik) \notin E} } \cos(2 \gamma J_{jk}) \times \\
      & \big[\ \cos(2\gamma (J_{ii} + J_{jj})) \prod_{\substack{(ik) \in E \\ (jk) \in E} } \cos(2 \gamma (J_{ik} + J_{jk} ) ) - \cos( 2 \gamma(J_{ii} - J_{jj})) \prod_{\substack{(ik) \in E \\ (jk) \in E} } \cos(2 \gamma (J_{ik} - J_{jk})) \big],\
\end{align*}

\section{\\Table for A.R. results for the Decomposition QAOA and Decomposition Gurobi Algorithms}\label{sec:appendixresults}

\begin{center}
\begin{tabular}{ |p{2.5cm}|p{2.5cm}|p{2.5cm}|p{2.5cm}| }
\hline
\multicolumn{4}{ |c| }{Approximation ratio} \\
\hline
QAOA & Decomp QAOA & Decomp Gurobi & H1-1 Results \\ \hline
0.75 & 0.96 &  0.96 & 0.99 \\ 
0.77 & 0.97 &  0.98 & \\
0.75 & 0.97 &  0.97 & \\ 
0.76 & 0.96 &  0.98 & \\ 
0.76 & 0.96 &  0.96 & \\
0.76 & 0.95 &  0.94 & \\
0.76 & 0.96 &  0.95 & 0.96 \\ 
0.77 & 0.97 &  0.97 & \\
0.76 & 0.98 &  0.96 & 0.93 \\ 
0.77 & 0.96 &  0.96 & \\
0.76 & 0.96 &  0.97 & \\ 
0.76 & 0.97 &  0.95 & 0.95\\ 
0.75 & 0.93 &  0.94 & \\
0.76 & 0.94 &  0.96 & 0.97 \\ 
0.76 & 0.96 &  0.96 & \\ 
0.74 & 0.97 &  0.95 & \\
0.75 & 0.94 &  0.95 & \\ 
0.76 & 0.96 &  0.95 & 0.96 \\ 
0.77 & 0.97 &  0.97 & 0.97 \\ 
0.76 & 0.95 &  0.95 & 0.98\\ 
0.75 & 0.95 &  0.96 & 0.98\\ 
0.76 & 0.95 &  0.96 & \\
0.77 & 0.95 &  0.98 & \\
0.76 & 0.94 &  0.97 & \\
0.77 & 0.98 &  0.97 & 0.99 \\ 
\hline
\end{tabular}
\captionof{table}{Approximation ratios, from left to right: Simulated QAOA, simulated decomposition algorithm with QAOA solving the decomposition, simulated decomposition algorithm with Gurobi solving the decomposition, hardware test on H1-1. Only 10 randomly selected graphs were tested on H1-1: those not tested have a blank entry.}\label{tab:results}
\end{center}


\end{document}